\documentstyle[pre,aps,epsf]{revtex}
\begin{document}
\draft
\title{
{\vspace{-2cm} \normalsize
\hfill \parbox[t]{4cm}{CERN-TH/96-323\\
                    ITP-Budapest 522\\}\\[3em]}
       Determining the $\beta$-function of the strong\\
                   interaction and closing the light gluino window}
\author{F. Csikor$^{(1)}$ and Z. Fodor$^{(2)}$\cite{home}}
\address{
$^{(1)}$ Institute for Theoretical Physics, E\"otv\"os University,
H-1088 Budapest, Hungary\\
$^{(2)}$ Theory Division, CERN, CH-1211 Geneva 23, Switzerland}

\maketitle
\begin{abstract} \noindent
We discuss the perturbative running of $\alpha_s$ in a model 
independent way. Our analysis contains data on the hadronic $\tau$ decay
and  hadronic cross sections in $e^+e^-$ annihilation between 
5 GeV and $M_Z$.  
We determine the color coefficients and the perturbative $\beta$-function  
of the strong interaction.  
 The results are in agreement 
with QCD and rule out the QCD+light gluino scenarios on the 70.8 -- 93.0\%  
CL. We combine our method with that of the multi-jet analysis at LEP. 
The combined result rules out light gluinos at least on the 99.76\% CL, 
provided that nonperturbative effects are not large.
\end{abstract}

\vspace{.5cm}

Asymptotic freedom is one of the most interesting predictions of QCD.
In order to study the running of $\alpha_s$, one can            
 collect its values at different scales (e.g. $\tau$ 
decay, DIS, $\Upsilon$ decay, $e^+e^-$ ) and 
 compare them \cite{schmelling96}. 
Another way is  to analyse a single experiment 
(e.g. $\bar pp$, $ep$ or  $\tau$ decay \cite{neubert95}).   
Nice agreement has been found between experiments and QCD.
Some differences between the results obtained from
low-energy and high-energy experiments 
led to a number of speculations; e.g. that
the apparent slower running of $\alpha_s$ could be due to additional
light fermions of the theory. 

Supersymmetric phenomenology deals normally with sparticles of 
masses ${\cal O}(100 GeV)$. The only exception  is the 
light gluino \cite{farrar78} with mass  $\le 1.5$ GeV (window I) 
\cite{farrar96,coulter} and $3-5$ GeV 
(window III) \cite{farrar94}. We will discuss window I and window III 
separately.  Since such a light particle influences
the running of $\alpha_s$ the comparison of high and
low energy experiments could open or close the light gluino window
\cite{antoniadis91}--\cite{ellis93}. A consistent analysis
must contain the virtual gluino effects not only in the running 
(technically in the $\beta$-function),   
but in all loop diagrams, extracting $\alpha_s$ from the experimental 
results, too \cite{ellis93}.

One can even determine   the $\beta$-function of the
strong interaction as done in ref. \cite{neubert95}. In order to extract 
the three-loop coefficients one needed large values of $\alpha_s$, that
is the small energies of $\tau$ decay. The result is in good 
agreement with QCD. 

It has been suggested \cite{campbell82} to look for gluinos in
four-jet events. The LEP collaborations 
determined the color coefficients ($C_A/C_F$ and $T_F/C_F$) 
in multi-jet events. The results (c.f. \cite{schmelling96})
are in good agreement with QCD. An extension of QCD including
light gluinos would result in a somewhat different $T_F/C_F$.  
The four-jet cross-section is known only at tree-level.   
Without the knowledge of the loop corrections this method alone
can hardly give a clear answer to the question of light gluinos.  

In this letter we determine the color coefficients and 
the perturbative $\beta$-function of the
strong interaction from experimental results. 
We consider gauge theories with fermions in the fundamental
representation and 
their extended versions with light gluinos. We perform an 
analysis on the three-loop level in the 
${\overline {MS}}$ scheme. The experimental
inputs are $R_\tau$ and $R_{h}$ for energies between $5$ GeV 
and $M_Z$. We combine our 
method with that of the multi-jet analysis at LEP.

One of the clearest ways to study the running of the coupling, thus 
the $\beta$-function, of the theory would be to study one given
experimental quantity at different energies. 
We chose  $R_{h}$'s at energies larger than 5 GeV,
which are  very clear quantites both experimentally and theoretically, 
due to their minimal nonperturbative corrections. Moreover, these are 
the only quantities known to ${\cal O}(\alpha_s^3)$ relative to the 
leading order, thus only a minimal scale ambiguity is present.
In order to study the running of $\alpha_s$
one needs precise $R_{h}$ measurements for a large energy region.
Unfortunately, $R_{h}$ measurements are 
limited by their small statistics for energy scales below $M_Z$. 
For this reason we have also considered another, strongly related
(still not very low scale) 
quantity $R_\tau$ in our analysis. Since we are aware
of the theoretical criticism on $\alpha_s$ determinations based
on $R_\tau$ we include the effects of all known 
uncertainties in case of QCD \cite{altarelli95}--\cite{shifman95}.  
The nonperturbative estimates to $R_\tau$ in the light gluino case 
might be unreliable. Therefore, we do not include  $R_\tau$ in our 
analysis   for window I gluinos. (Window III gluinos are much too heavy to 
contribute to $R_\tau$.)

Let us suppose that the strong interaction is described by a
gauge theory based on a 
simple Lie-group, which fixes the color coefficients of the
theory ($C_F,C_A,T_F$). The three basic processes in the theory are:
gluon bremsstrahlung from a quark, splitting of a gluon into two gluons 
and splitting of a gluon into two quarks. To lowest order 
their amplitudes are proportional to a universal coupling and to
$C_F,\ C_A$ and $T_F$, respectively. E.g. for an $SU(N)$ gauge group
one has $C_F=(N^2-1)/(2N),\ C_A=N$ and $T_F=1/2$.
The $\beta$-function of the strong interaction, hadronic cross-sections 
and widths are calculated in terms of these coefficients and the active 
number of fermions. We will look for a set of parameters which
describes the experimental data most accurately and give the
corresponding confidence level (CL) regions. 
The outcome --that is the best
fit-- does not necessarily predict a meaningful theory. Nevertheless, it 
could tell the difference between QCD and QCD+light gluino scenario.   

The $\beta$-function: 
$d a(\mu)/ d\log\mu=-\beta_0 a^2-\beta_1 a^3-\beta_2 a^4 \ $ 
is known upto three-loop order.
We factorize out the inconvenient $\pi$-s by using   
$a=C_F\alpha_s/(2\pi)$. Introducing $x=C_A/C_F$ and $y=T_F/C_F$
gives 
$\beta_0=11x/3-4(n_f y+xn_{\tilde g}/2)/3$, where $n_f$ is the number 
of active flavours and $n_{\tilde g}$ is the number of active gluinos.
The expressions for $\beta_1$ and $\beta_2$ are more 
complicated \cite{gorishni91,clavelli355}. 
 In the running of $\alpha_s$ 
we follow ref. \cite{bernreuther82} for  
 threshold-effects. We solve the above renormalization group equations
 exactly.

The other important quantity in the analysis --the hadronic cross section in 
$e^+e^-$ annihilation via a virtual photon-- is known on the three-loop level, 
too.
\begin{eqnarray}
R_\gamma&=&\sigma(e^+e^-\rightarrow\gamma\rightarrow hadr)/ 
\sigma_{o}(e^+e^-\rightarrow \gamma \rightarrow \mu \mu )\\ \nonumber
&=&3 \Sigma_f q_f^2(1+K_1a+K_2a^2+K_3a^3+...), \nonumber 
\end{eqnarray}
where $K_1=3/2$.
$K_2$,  $K_3$
is more complicated and also known for arbitrary $C_F,\ C_A$ and $T_F$
\cite{gorishni91,clavelli96}. 
The hadronic decay ratio of the $\tau$ is defined:  
$R_\tau=\Gamma(\tau^-\rightarrow \nu_\tau+hadr.) /
\Gamma(\tau^-\rightarrow \nu_\tau e^- \bar\nu_e).$ 
Its perturbative value is strongly related to $R_\gamma$.
 For a recent analysis and a review on $R_\tau$ see eg. 
\cite{neubert96} and \cite{pich95}. 
The hadronic decay width of the $Z$ boson
$R_Z=\Gamma(Z\rightarrow hadr.)/
\Gamma(Z\rightarrow \mu^+\mu^-)  $
is again a function of $R_\gamma$ and known at the three-loop  level.
At the energy scales of the analysis important electroweak and mixed
corrections appear. We include these corrections.

The experimental values for $R_\tau$ are given in \cite{rtau} with an 
average of $R_\tau=3.616 \pm .02$. 
Averaging for the four LEP experiments and 
three leptons \cite{ewgroup} gives $R_Z=20.778 \pm .029$. The references for 
hadronic cross sections at
energies below $M_Z$ are taken from \cite{lowenergy}.
We have collected all the existing published data, and some
unpublished results, too. Some of them were binned by the experimental 
groups. The total number of data points included in
our analysis is 182.

Combining results from different experiments is a delicate question, 
with major problems. i) The use of the
radiative corrections is  not unique, reflecting the state of art
at the time of publication. 
 ii) The results do
depend on the mass of the Z boson and top-quark --which was assumed
to be approximately 20 GeV at the early eighties. iii) The measurement
of the total cross section was not performed in the
full phase space, and Monte-Carlo acceptance
calculations based on different assumptions were done. 
We corrected for the first two problems, 
correction for the third one is practically impossible. For
further details see \cite{cello87}--\cite{agostini89}.    

There are different sources of uncertainties in the determination
of $\alpha_s$. We treat them in a unified manner.
 We add the systematic errors linearly. The total systematic 
error estimates and the statistical errors are combined quadratically. 
The overall normalization errors within one experiment are 
correlated. We include these correlations. 
We minimize   
$\chi^2=\Delta^TV^{-1}\Delta$, where $\Delta$ is an $n$-vector of 
the residuals of $R_i-R_{fit}$ for the $n$ individual  
results and V is an $n\times n$ error matrix. In $V$, the diagonal 
elements $V_{ii}$ are the squares of the total errors for the $i$th
measurement and the off diagonal elements $V_{ij}$ correspond
to the correlations between the $i$th and $j$th measurements.
For different points from the same experiment $V_{ij}$ is
given by the product of the normalization errors \cite{cello87}. 
The separation of the systematics into point-to-point 
and overall normalization error is 
ambiguous. Our results are rather
insensitive to a moderate change of this splitting in a given 
experiment. For some experiments
the above separation was not even explicitly given. 
We checked that our results are stable against even large variations
of the splitting in these cases. However, the unreasonable
extreme case of totally uncorrelated fitting would give rather 
different results with much more predictive power.  
We have assumed that the results of different experiments are
uncorrelated. In order to perform a further consistency check we 
included an additional hypothetical overall error of  1\%
in the correlation matrix.
The change of the result turned out to be negligible.   
Since the experimental groups usually give their results 
with binning in fixed energy intervals we checked that
further binning had practically no influence on our results.

\begin{figure}                                                                    
\centerline{\epsfxsize=0.8  \linewidth \epsfbox{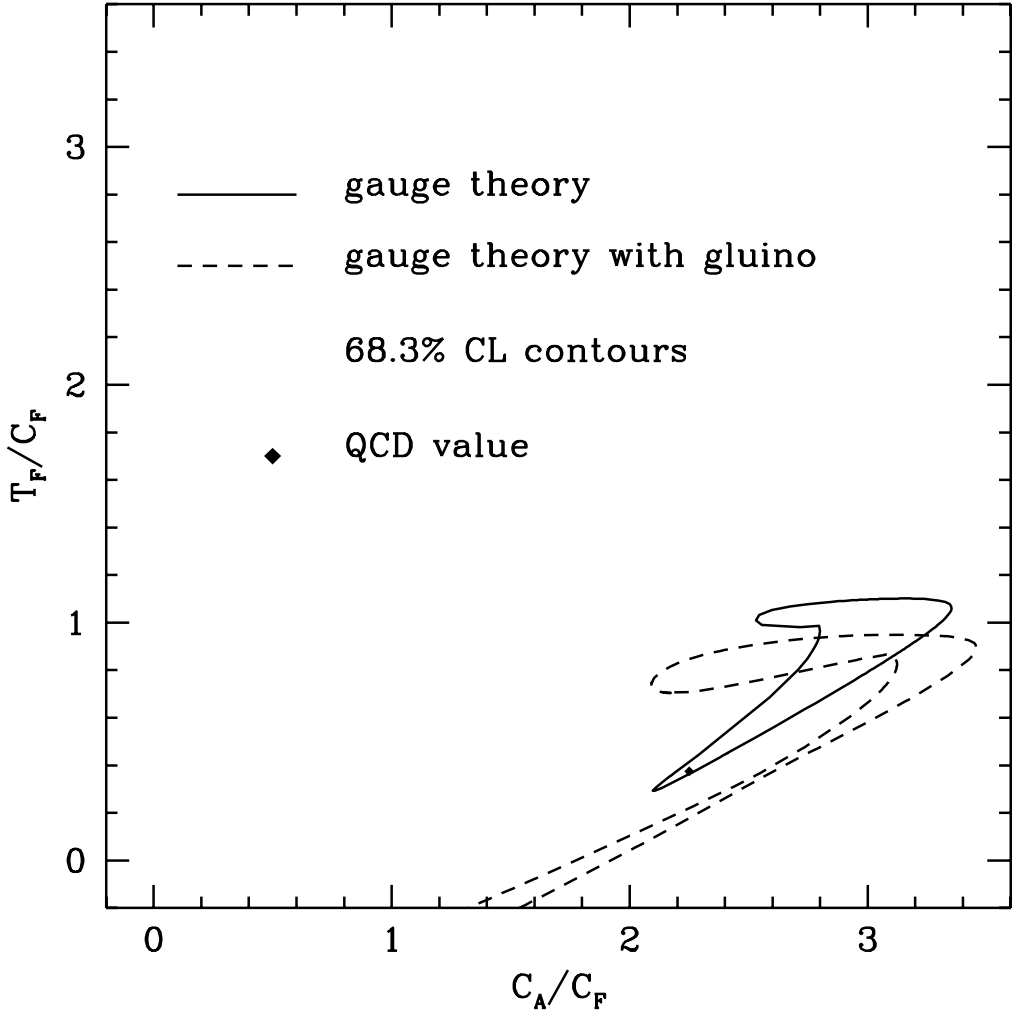}}
\caption{\label{fig1}
{The 68.3\% CL regions for $x=C_A/C_F$ and $y=T_F/C_F$.
 The QCD value is represented by a square.
}}
\end{figure}

Due to the relatively poor statistics for
the $e^+e^-$ experiments below $M_Z$ the theoretical uncertainties 
are dominant only for $R_\tau$ and non-negligible for$R_Z$. 
Estimating the nonperturbative and perturbative errors in 
$\tau$-decay we use the results of \cite{altarelli95,neubert96}.
For a fixed order QCD calculation we assume  
that the error is equal to the last computed term.  
Including all the errors we get $R_\tau =3.616 \pm .143$, which corresponds 
to $\alpha_s (M_\tau)=.335 \pm .053$ and $R_Z =20.778 \pm .0387$, 
which corresponds to $\alpha_s (M_Z)=.123 \pm .006$.  
We assume that currently incalculable and/or model dependent nonperturbative
corrections are negligible or correctly estimated. If the nonperturbative 
corrections turn out to be larger than our estimates, the confidence levels 
for light gluino exclusion to be determined below should be lowered. Note 
that both error estimates are certainly very conservative.

Having included all the errors one can determine the best fit values
and the CL regions for $x=C_A/C_F$ and $y=T_F/C_F$ with or 
without light gluinos.

Fig. 1 contains the 68.3\% CL regions
for theories without gluinos and with window III  gluinos. 
As it can be seen
the experimental results are in agreement with QCD; however, they
rule out the QCD+light gluino scenario on the 93 (90.7)\% 
CL's for
$M_{\tilde g}=3\ (5)$  GeV, respectively. 
The variation of the fixed $\chi^2 $ boundaries is quite small 
for the above mass region.  
For binning level 50 (d.o.f. 48) the $\chi^2$ values are 
37.37 without and 37.94 with gluinos 
($M_{\tilde g}=3$ GeV) for the best fits. 
As mentioned before, for window I gluinos we do not include $R_\tau $ in our 
study. In this case we have a 70.8\% exclusion only.

\begin{figure}                                                                    
\centerline{\epsfxsize=0.8  \linewidth \epsfbox{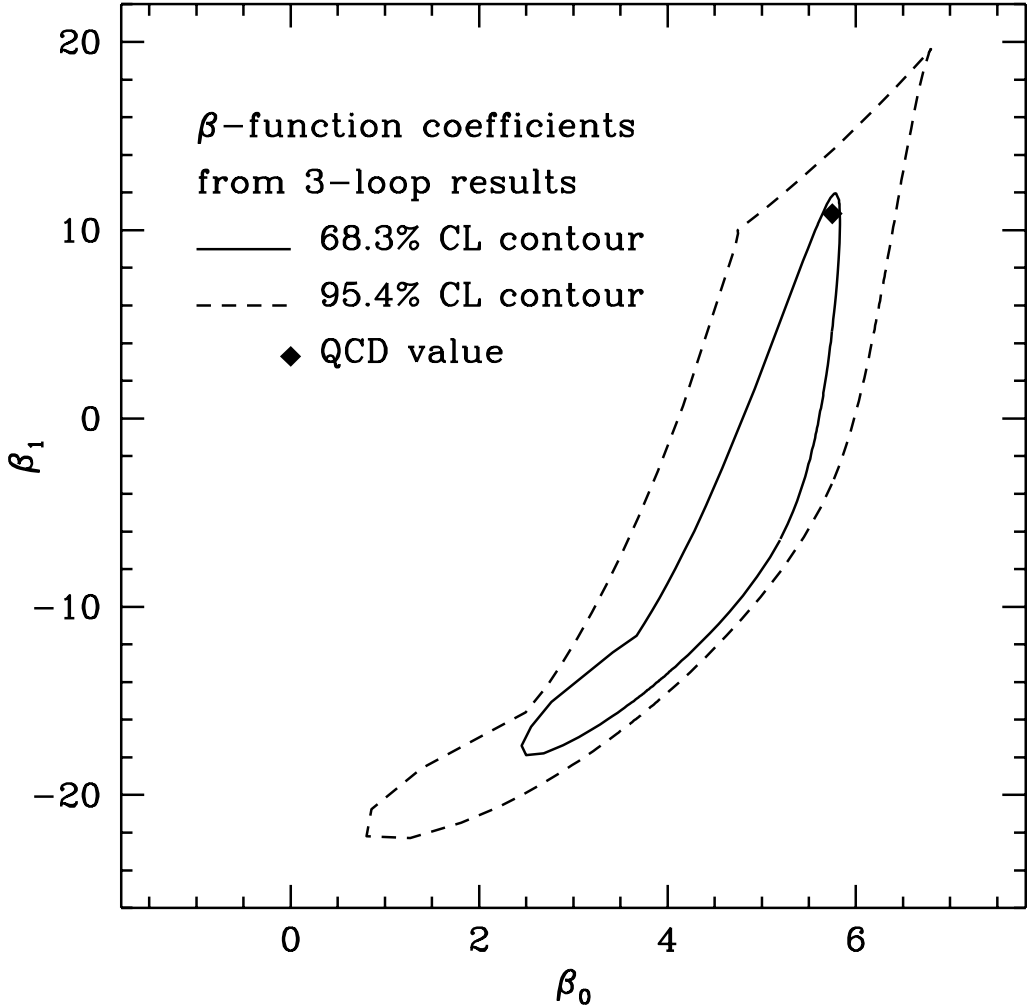}}
\caption{\label{fig2}
{The 68.3\% and 95.4\% CL regions for $\beta_0$
and $\beta_1$.
}}
\end{figure}

Note the important 
difference between the presentations of our result and that
of the 4-jet analysis for the case with light gluinos \cite{schmelling96}.
In our case the best fits and the CL contours
for the $x,y$ variables are determined separately for the 
theories with and without light gluinos. These results
are compared for both theories with the QCD values: 
${\bar x} =2.25,\ {\bar y} =.375$. Due to the simplicity of the
tree-level treatment of the 4-jet analysis in ref. 
\cite{schmelling96}, it was possible
to produce the same best fits and contours for both theories
and compare this unique result with ${\bar x} ,{\bar y} $ above 
in the theory without light gluinos, and
with ${\bar x} $ and with an effective ${\bar y}_{eff}=.6$ 
for the gluino extended theory.
The 4-jet results can easily be expressed
in terms of our variables.

Two remarks are in order: 
i) the 4-jet analysis is based on tree-level calculations, whereas our
method contains corrections up to three-loops. (In 
\cite{schmelling96} also the one-loop analysis to 2- and 3-jets
is done; however, the errors on $x,y$ are huge.)
ii) our method contains the running of  
$\alpha_s$, too, thus it gives a simultaneous check of the 
$\beta$-function. 

Assuming \cite{murayama} that the underlying gauge group is SU(3), we may 
consider the number of gluinos as a free parameter. The best fits give 
$n_{\tilde g} = -.638\pm 1.17$
for  window I and $n_{\tilde g} =.0078\pm .52$ for window III.
The CL's are given by Bayes' theorem allowing only non-negative integer 
$n_{\tilde g}$: 72.5\% for  window I and 87.7\% for window III.
(For modeling data with bounded physical region: $n_{\tilde g} \geq 0$, 
and for application of Bayes' theorem see e.g. Sec. 28 of \cite{lowenergy}.) 
Finally, one can also fix the number of gluinos and determine 
$\alpha_s (M_Z )$. For $n_{\tilde g} =0$ (i.e. QCD) we get $.124 \pm .004$, 
for $n_{\tilde g} =1$ we get $.129 \pm .006$ (window I) and 
$.132 \pm .006$ (window III). As expected \cite{lowenergy} the QCD value is 
slightly  larger than  the world average.

\begin{figure}                                                                    
\centerline{\epsfxsize=0.8  \linewidth \epsfbox{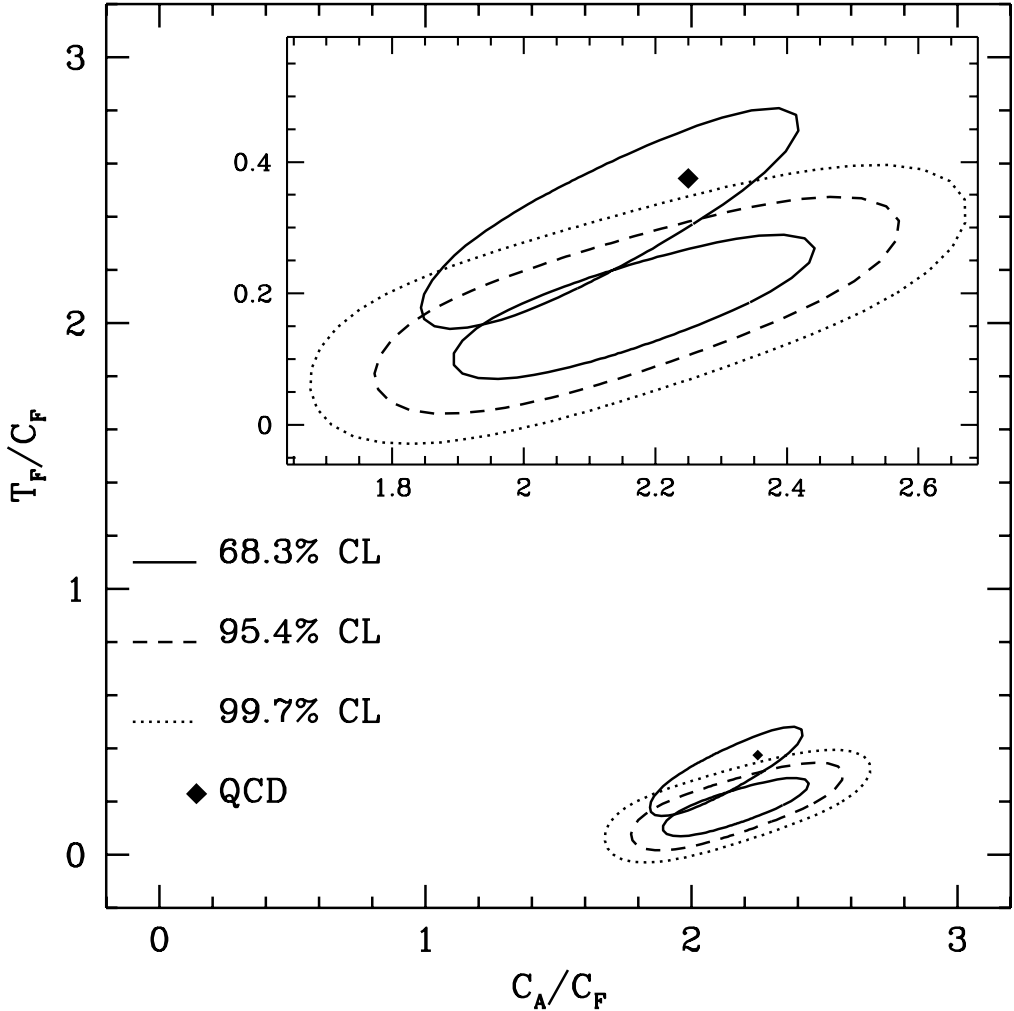}}
\caption{\label{fig3}
{The 68.3\%, 95.4\% and 99.7\% CL regions of the combined analysis 
for the theory with window III light
gluino ($M_{\tilde g} =3$ GeV) and the 68.3\% CL region for the theory without 
gluino.
}}
\end{figure}

Fig. 2 shows the $\beta$-function coefficients 
with 68.3\%  and 95.4\% CL regions 
for models without gluinos at $M_Z$. 
These curves represent merely a transformation of the
curves in the $(x,y)$ plane, they result in similar
conclusions for QCD. As it can be seen the value
of $\beta_0$ is quite well constrained;  
however, not even
the sign of $\beta_1$ could be predicted. This result
reflects the fact that in the energy region studied 
the running of the coupling is determined
almost exclusively by the one-loop term, proportional to $\beta_0$.
An interesting possibility  is to fix the cross sections,
$\beta_1$, $\beta_2$ and the flavour dependence of $\beta_0 $ to 
their perturbative QCD values and
extract $\beta_0$ from the fits. The result is $\beta_0=5.70 \pm .64$
to be compared with the theoretical value of 5.75.
Following \cite{neubert95} we fixed $\beta_0$ and $\beta_1$
to their theoretical (QCD) value and performed a fit with
$\beta_2$ as a free parameter. The allowed region for $\beta_2$
contains the theoretical prediction, but the error is 
an order of magnitude larger than the theoretical value.

Comparing our Fig. 1 and Fig. 10  of \cite{schmelling96} --taking proper
care   
of using the same variables-- one observes that the overlap of the
68.3\% CL regions for the two analyses is quite small. This
opens the possibility for a much stronger restriction on the theories
without and with gluino. We have parametrized the $\chi ^2 $ of the
multi-jet analysis as a function
of $x$ and $y$, and performed a new fit adding it to our $\chi ^2 $ function. 
In the actual analysis we have included the new results in \cite{aleph}. 
 Since the 4-jet analysis is based on 
a tree-level calculation we have included the unknown higher order QCD
and mass effects. 
We assumed much larger uncalculated higher order corrections 
than used in the experimental papers. Due to them we enlarged the axes of the error ellipse by  12\%  of the theoretical x and y values (relative correction of ${\cal O}(\alpha_s )$.)
For the 4-jet analyses we have used the mass
effects calculated by \cite{munoz94}. 
The results of the combined analysis for QCD and window III gluinos are presented in Fig. 3. 
Our result is
consistent with QCD + no gluino scenario, while the theory with window III
light
gluino is excluded on the 99.99 (99.89)\% CL for 
$M_{\tilde g}=3\ (5)$ GeV.
The window I gluino is excluded on the 99.97\% CL.

Again one can fix the underlying gauge group to SU(3) 
and consider the number of gluinos as a parameter to be fitted. We get 
 $n_{\tilde g}=-.35     \pm .33$ for window I and  
$n_{\tilde g}=-.156 \pm .27$
 ($n_{\tilde g}=-.197 \pm .32$) for window III $M_{\tilde g}=3\ (5)$ GeV 
gluinos.                 
Using Bayes' theorem assuming that  $n_{\tilde g}$ 
  is non-negative integer number, 
the exclusion CL's are  99.96\% for window I and 99.96 (99.76)\%  for 
 window III gluinos.

Leaving out $R_\tau $ from the window III combined analyses, 
the CL values change only negligibly.

In summary, we have presented a method to analyse the running of $\alpha_s$ 
 from $M_\tau$ to $M_Z$. We have determined the group 
coefficients of the theory, which are in agreement with QCD.
The $\beta$-function coefficients are also given with a
similar result. Furthermore, our results rule out
the QCD+light gluino scenario on the 70.8\% and 90.7--93.0\% CL for 
window I and III, respectively. 
Combining the analysis of the present letter and the method
using multi-jet data (cf. ref. \cite{schmelling96,aleph}, however 
treating the theoretical errors more conservatively than the
experimental groups)
we have got an order of magnitude smaller CL 
regions for the allowed group coefficients, which supports QCD within 
$1\sigma$ and excludes the 
QCD+light gluino scenario with a 99.97\% for window I and at least 99.89\% CL 
 for window  III gluinos. 
Performing a one parameter fit to determine the number of gluinos, we 
get $n_{\tilde g}=-.35 \pm .33$ and 
$n_{\tilde g}=-.197 \pm .32$  for window I
 and III, respectively. In terms of CL's this corresponds to 99.96\% 
and 99.76\%.
Needless to say, our results are valid assuming the light gluino extension 
of the standard model. We have not investigated whether additional 
new physics could change our exlusion CL limits.

The details of the above analysis, the convergence features and the 
statistical analysis will be presented in a forthcoming 
publication \cite{alom}.

We thank M. Drees, B. Gary, A.L. Kataev and R. Sommer for discussions. 
This work was partially supported by Hung. Sci. 
grants OTKA-T16248,T22929.

\vfill
\end{document}